\documentclass[useAMS,usenatbib]{mn2e}

\usepackage{amsmath,amssymb,wasysym,psfig,graphicx,lscape,natbib,epsfig}
\bibpunct{(}{)}{;}{a}{}{,}

\def\mn={MNRAS}

\title[Short GRBs from SGR flares and NS mergers]{Short gamma-ray bursts from SGR giant flares and neutron star mergers: two populations are better than one}
\author[Chapman R. and others]{Robert Chapman$^1$\thanks{Email:r.1.chapman@herts.ac.uk}, Robert S. Priddey$^1$ and Nial R. Tanvir$^2$\\
$^1$Centre for Astrophysics Research, University of Hertfordshire, College Lane, Hatfield AL10~9AB, UK\\
$^2$Department of Physics and Astronomy, University of Leicester, Leicester, LE1~7RH, UK}

\begin{document}

\date{6 January 2009}


\maketitle

\label{firstpage}

\begin{abstract}
There is increasing evidence of a local population of short duration Gamma-ray Bursts (sGRB), but it remains to be seen whether this is a separate population to higher redshift bursts. Here we choose plausible Luminosity Functions (LF) for both neutron star binary mergers and giant flares from Soft Gamma Repeaters (SGR), and combined with theoretical and observed Galactic intrinsic rates we examine whether a single progenitor model can reproduce both the overall BATSE sGRB number counts and a local population, or whether a dual progenitor population is required. Though there are large uncertainties in the intrinsic rates, we find that at least a bimodal LF consisting of lower and higher luminosity populations is required to reproduce both the overall BATSE sGRB number counts and a local burst distribution. Furthermore, the best fit parameters of the lower luminosity population agree well with the known properties of SGR giant flares, and the predicted numbers are sufficient to account for previous estimates of the local sGRB population.
\end{abstract}


\section{Introduction}

Results from the Burst and Transient Source Experiment (BATSE) onboard the \textit{Compton Gamma-Ray Observatory} showed that Gamma-ray Bursts (GRBs) divide observationally into two classes based primarily on their duration~\citep{1993ApJ...413L.101K}: long GRBs have durations $>$ 2 seconds, and short GRBs $\le$ 2 seconds. Short GRBs (sGRBs) seem to be associated with a variety of host galaxies with no apparent restriction on galactic properties~\citep{2006ApJ...642..989P,2007ApJ...670.1254B,2008MNRAS.384..541L}, although host identification is not always trivial~\citep{2007MNRAS.378.1439L}. Additionally, a handful of recently detected sGRBs have localisations consistent with origins in nearby galaxies~\citep{2006ApJ...652..507O,2007ApJ...662.1129O,2007AstL...33...19F,2007arXiv0712.1502M,2008MNRAS.384..541L}. Overall, the \textit{Swift} redshift distribution of sGRBs~\citep{2007ApJ...670.1254B} peaks closer than that of long GRBs~\citep{2006A&A...447..897J}, though there is evidence that some sGRBs may occur at higher redshifts~\citep{2006ApJ...648L...9L}, and that there may be a local population of underluminous long GRBs~\citep[e.g.][]{2006Natur.442.1011P,2006Natur.442.1014S,2006ApJ...638L..67L,2007ApJ...662.1111L,2007MNRAS.382L..21C}.

The leading progenitor model for sGRBs is the merger of two compact objects, neutron star-neutron star (NS-NS) or neutron star-black hole~\citep{2007PhR...442..166N} binaries. The Luminosity Function (LF) of BATSE sGRBs has been investigated previously assuming a single progenitor population~\citep[e.g.][]{2001ApJ...559L..79S,2004JCAP...06..007A,2005A&A...435..421G,2006A&A...453..823G,2006ApJ...643L..91H} in order to determine the intrinsic rate and most likely LF parameters. In a refinement to their previous work, \citet{2006A&A...453..823G} noted that a second population of bursts may be necessary to explain some features of their model fits and the comparison with \textit{Swift} bursts, particularly at lower redshifts. \citet{2006ApJ...643L..91H} considered both primordial and dynamically formed NS binaries, and suggested that the early observed redshift distribution of sGRBs favoured dynamical formation. Further to that work, \citet{2007arXiv0710.3099S} suggested that the more recent \textit{Swift} cumulative redshift distribution is better encompassed by including both formation routes with different abundances above and below $z\sim0.3$. Recently, in an analysis of a large number of models of compact object merger scenarios from population synthesis models, \citet{2008ApJ...675..566O} have shown that the observed sGRB redshift distribution could be reproduced by a reasonable fraction of those models, though this analysis was insensitive to the low end of the redshift distribution on which our work here is focused. \citet{2006ApJ...650..281N} find the high rate of observed sGRBs within 1~Gpc to imply that a single population of NS binaries responsible for all sGRBs must be dominated by long merger times, inconsistent with the observed NS binary population. However, they also point out that a non-unimodal luminosity function, such as produced by two separate populations of progenitor, cannot be ruled out for sGRBs.


There are indeed other possible progenitors for sGRBs. At much closer distances still, the initial spike in a giant flare from a Soft Gamma Repeater (SGR) in a relatively nearby galaxy would also appear as a sGRB. For example, the December 27th 2004 event from SGR1806-20 would have been visible by BATSE out to $\approx50~\rm{Mpc}$~\citep{2005Natur.434.1098H,2005Natur.434.1107P,2006MPLA...21.2171T,2007PhR...442..166N}. Thus it is entirely plausible that some fraction of sGRBs are extragalactic SGR giant flares. Several studies have estimated the likely contributions of SGR flares to BATSE sGRBs. \citet{2006MNRAS.365..885P} estimate a rate of a few percent based on a lack of definite sGRB detections among the shortest BATSE GRBs consistent with locations within the Virgo cluster. Searches for hosts plausibly connected with six well-localised sGRBs~\citep{2006ApJ...640..849N} suggest a rate of less than $15\%$, and in a comparison of the spectra of a limited sample of the brightest BATSE sGRBs, \citet{2005MNRAS.362L...8L} conclude only $4\%$. \citet{2005Natur.434.1107P}, based again on a lack of events from the Virgo cluster, find a rate of less than $5\%$, though point out that the LF of SGR giant flares may extend to much larger luminosities, such as suggested by~\citet{2002MNRAS.335..883E}. \citet{2007ApJ...659..339O} points out that the fraction cannot be less than $\sim1\%$ without being inconsistent with the observed Galactic SGR giant flare rate, and calculate an upper limit of $16\%$ ($95\%$ confidence limits) based on a conservative measure of probable IPN sGRB coincidences with bright star forming galaxies within 20~Mpc. This limit is sensitive to their estimate of the completion of the galaxy sample and may be higher still.

Previously, using the full sample of BATSE sGRBs localised to better than 10 degrees, we demonstrated that between 6 and 12 per cent of BATSE sGRBs were correlated on the sky with galaxies within $\approx28~\rm{Mpc}$~\citep{2005Natur.438..991T}, and we have now extended this work out to $\approx155~\rm{Mpc}$. Our analysis was based purely on burst/galaxy distribution correlations and unbiased with regards to burst brightness or other assumptions, though our later work showed that this correlation is dependent mostly on large scale structure on the sky rather than individual burst/galaxy pairings~\citep{2007MNRAS.382L..21C}. The main question we address in this paper is whether a nearby population ($z\le0.03$) of this magnitude may be produced by a suitable LF describing a single progenitor population, or whether it is necessary to include an intrinsically lower luminosity population as well.

Here we attempt to answer this question by considering first single, and then dual population LFs. The intrinsic rates in the models will be assumed from both the observed Galactic SGR flare rates and the modelled NS-NS merger rates in order to investigate the LF parameters. Obviously there are significant uncertainties in these rates: the Galactic giant flare rate in particular is estimated from only 3 observed events. Regardless of these uncertainties and the exact form of luminosity functions chosen, we find that a single progenitor population described by a unimodal (i.e. with a single peak or knee) LF cannot produce sufficient local events, whereas a dual population reproduces the likely local sGRB distribution as well as the overall number counts\footnote{Throughout this paper we assume a flat cosmology with $\rm{H_0}=71\rm{km~s^{-1}~Mpc^{-1}}$, $\Omega_M=0.27$ and $\Omega_\Lambda=0.73$}.

\section{Methods}
The number of sGRBs, $N$, observed above a threshold $p$ in time $T$ and solid angle $\Omega$ is given by Equation~\ref{lfeq}, where $\Phi(L)$ is the sGRB LF, $R_{GRB}(z)$ is the comoving event rate density at redshift $z$, $dV(z)/dz$ is the comoving volume element at $z$ and $z_{max}$ for a burst of luminosity $L$ is determined by the detector flux threshold and the luminosity distance of the event.

\begin{equation}
N(>p)=\frac{\Omega T}{4\pi}\int_{L_{min}}^{L_{max}}\Phi(L)dL\int_{0}^{z_{max}}\frac{R_{GRB}(z)}{1+z}\frac{dV(z)}{dz}dz \label{lfeq}
\end{equation}

We are of course dealing with detector limited and not bolometric luminosities. Following \citet{2001ApJ...559L..79S} and \citet{2005A&A...435..421G} we assume a constant median spectral index of $-1.1$ in the BATSE energy range of 50-300~keV to derive a simplified K correction and conversion to photon flux.

\subsection{Intrinsic rates}
The sGRB rate per unit volume, $R_{\textrm{GRB}(z)}$ is given by Equation~\ref{rgrb} where $ N_{\textrm{GRB}}$ is the number of sGRBs per progenitor, $\rho_{\textrm{progenitor}}$ is the intrinsic ($z=0$) progenitor formation rate and $F(z)$ describes the volume evolution of this rate with $z$. 

\begin{equation}
R_{\textrm{GRB}}(z)= N_{\textrm{GRB}} \times \rho_{\textrm{progenitor}} \times F(z)~\textrm{Mpc}^{-3} \label{rgrb}
\end{equation}

For NS-NS mergers, a burst is produced only once at merger, and we therefore assume $ N_{\textrm{GRB}}=1$. This is of course an upper limit: any beaming of sGRBs, or a GRB production efficiency per merger of less than $100\%$, would effectively reduce this number, and reduce the number of bursts observable from the NS-NS merger population. This limit is therefore conservative in the sense that it maximises the possible fraction of bursts produced by mergers in our analysis. The intrinsic NS-NS merger rate is taken as $10^{-5}~\rm{yr}^{-1}$ per Milky Way equivalent galaxy (Star Formation Rate, SFR $\approx4\rm{M_\odot~yr^{-1}}$, e.g.~\citet{2006Natur.439...45D}) from the population synthesis models of \citet{2007PhR...442...75K}. Mergers, of course, occur some time after the formation of the binary itself. Thus the merger rate at redshift $z$, is dependent not on the SFR at the same $z$, but on the earlier SFR at higher redshift. $F(z)$ is therefore given by the convolution of the SFR as a function of redshift with a distribution of delay times from binary formation to merger. The population syntheses of \citet{2006ApJ...648.1110B} suggest a merger delay time (formation plus coalescence) distribution $dP_{m}/d(\log(t)) \sim constant$ ($\equiv dP_{m}/dt \propto 1/t$) between $10^7$ and $10^{10}$ years, with a narrow peak at the very lowest times, and we thus assume a delay time probability distribution where $dP_{m}/d(\log(t))$ is flat between $10^7$ and $10^{10}$ years and zero outside this range, for simplicity and comparison with previous LF analyses. We note, however, that using a delay model including a narrow early `spike' (with an order of magnitude higher value between 15 and 30~Myr) makes little difference to the derived LF parameters as can be seen from some examples in Table~\ref{results_single}.

SFR as a function of $z$ is parameterised according to the SF2 model of \citet{2001ApJ...548..522P}, normalised to a local SFR of $1.3\times10^{-2}~\rm{M_{\odot}yr^{-1}Mpc^{-3}}$~\citep{1995ApJ...455L...1G} as given in Equation~\ref{SFR}.

\begin{equation}
\textrm{SFR}(z)=1.3\times10^{-2}\left(\frac{23e^{3.4z}}{e^{3.4z}+22}\right)~\textrm{M}_{\odot}~\textrm{yr}^{-1}~\textrm{Mpc}^{-3}\label{SFR}
\end{equation}

An alternative analysis is that the merger rate should be proportional to Stellar Mass Density (SMD), which must be representative of star formation history. We therefore also investigate merger rates which follow a simple single exponential fit to the SMD out to $z\sim5$ derived from the FORS deep field~\citep{2005ApJ...619L.131D} as:

\begin{equation}
\textrm{SMD}(z)=10^{8.75}\exp(-\ln(2)z)~\textrm{M}_{\odot}~\textrm{Mpc}^{-3}\label{mass}
\end{equation}

Over the last 30 years of observations, there have been 3 giant flares from 4 known SGRs in the Milky Way and Magellanic Clouds. The observed local rate of giant flares per Galactic SGR is therefore $\approx3\times10^{-2}~\rm{yr^{-1}}$, and their short active lifetimes of $\sim10^4$ years~\citep{1992ApJ...392L...9D,1998Natur.393..235K,1999PNAS...96.5351K} imply $ N_{\textrm{GRB}}\sim300$ in the SGR case. Magnetars are commonly believed to form in a fraction of core collapse supernovae, and hence their formation should follow the SFR as a function of $z$. Given the the association of the 4 known SGRs with young stellar populations, this therefore implies a formation rate via core collapse supernovae of $4\times10^{-4}~\rm{yr^{-1}}$. 

However, it is also plausible that magnetars may form via the Accretion Induced Collapse (AIC) of White Dwarf (WD) binaries which contain at least one sufficiently massive and magnetized member~\citep{2006MNRAS.368L...1L}. In older galaxies with relatively little star formation, this would be the dominant formation route and therefore makes it possible for SGRs to be associated with all types of galaxies, not just those with a relatively high SFR. Following \citet{2006MNRAS.368L...1L}, the rate of magnetar formation via WD-WD mergers in a Milky Way equivalent galaxy is estimated as $3\times10^{-4}~\rm{yr^{-1}}$. We therefore assume $F(z)$ for SGRs follows both SFR(z) for magnetar production from supernovae and either the delayed SFR or SMD to allow for production by WD binary mergers.

\subsection{Luminosity functions}

Luminosity functions for SGR giant flares and NS-NS mergers are not well known. A lognormal LF approximates the shape of the theoretical NS-NS merger luminosity distribution~\citep{2003MNRAS.343L..36R}, but other functional forms may be equally valid: for example \citet{2005A&A...435..421G} assumed a broken power law for their LF calculations, and the luminosities of many other astronomical populations are well described by a Schechter function~\citep{1976ApJ...203..297S}.

Given only 3 events, it is not possible to constrain the SGR giant flare LF to any great degree. The 3 observed events have peak luminosities of $\sim10^{44},~\sim10^{46}$ and $\sim10^{47}~\rm{erg~s^{-1}}$~\citep{2007ApJ...665L..55T} (including a correction for the lower distance estimate of SGR\,1806-20 found by \citet{2008MNRAS.tmpL..34B}). The more common short duration bursts from SGRs, with luminosities up to $10^{41}~\rm{erg~s^{-1}}$ seem to follow a power law distribution in energy, $dN \propto E^{-\gamma} dE$ where $\gamma\sim1.4-1.8$~\citep{1996Natur.382..518C,2000ApJ...532L.121G} similar to that found in earthquakes and solar flares. Intermediate bursts with energies and luminosities between the short bursts and giant flares are also seen, and it is possible therefore that this distribution continues to higher energies and includes the giant flares themselves, particularly since \citet{2000ApJ...532L.121G} found no evidence for a high energy cutoff in their work. However, \citet{1996Natur.382..518C} did find evidence of a cutoff around $5\times10^{41}$~erg, and furthermore the intermediate bursts are generally seen following giant flares and may be some form of aftershock rather than representing part of a continuous spectrum of flare activity. Theory suggests that the common bursts are produced by the release of magnetic energy gated by a small scale fracturing of the crust sufficient only to relieve crustal stresses, whereas the giant flares are the result of large scale cracking sufficient to allow external field reconfiguration to a new equilibrium state~\citep{1993ApJ...408..194T,1995MNRAS.275..255T}. Assuming the latter is a physically distinct process discontinuous (in terms of energy release) from the short bursts, then it must have some minimum energy release, and a maximum defined by the total destruction of the external field via the Flowers-Ruderman instability~\citep{1977ApJ...215..302F} where entire hemispheres of the magnetar flip with respect to each other~\citep{2002MNRAS.335..883E}. Having only the 3 observed events to go on, a lognormal LF is once again plausible for giant flare luminosities. The possibility of a continuous luminosity distribution between the short, intermediate and giant flares is not ruled out however, and we therefore also consider a single power law LF as well.

To summarise, we consider the possibility that short GRBs may be produced via two different progenitor routes, both NS-NS mergers and SGR giant flares, each population with intrinsically different luminosities. The forms chosen for the luminosity functions examined are as follows:

\noindent
1. Lognormal distribution 

\begin{equation}
\frac{dN}{d\log{L}} \propto \exp{\left(\frac{-(\log{L}-\log{L_0})^2}{2{\sigma}^2}\right)}
\end{equation}

\noindent
2. Schechter function

\begin{equation}
\frac{dN}{dL} \propto \left(\frac{L}{L_0}\right)^{-\alpha}
\exp{(-L/L_0)}~,~L \ge L_{min}
\end{equation}

\noindent
3. Power Law

\begin{equation}
\frac{dN}{dL} \propto \left(\frac{L}{L_0}\right)^{-\alpha}~,~L_{min} \le L \le L_0
\end{equation}

where $L_{min}=10^{42}~\rm{erg~s}^{-1}$ for normalisation and convergence of the Schechter function (see Appendix A for discussion of the limited effect of the choice of $L_{min}$). The Power Law distribution is normalised to the observed Galactic rate between $L_{GFmin}=10^{44}~\rm{erg~s}^{-1}$ and $L_0$, but the distribution is analysed down to $L_{min}$ to investigate the possible extension of the power law to lower luminosity flares. $L_0$, and $\alpha$ or $\sigma$ are the free parameters to be estimated.

\subsection{Constraining the models}

The $C_{max}/C_{min}$ table from the current BATSE catalogue~\citep{1999ApJS..122..465P} provides peak count rate for bursts in units of the threshold count rate. Not all bursts are included and in addition the BATSE threshold was varied historically. Therefore in order to analyse a consistent set of bursts we restricted the table to only those sGRBs recorded when the 64ms timescale threshold was set to $5.5\sigma$ above background in at least 2 detectors in the $50-300~\rm{keV}$ range. The all sky equivalent period (including correction for BATSE's sky coverage) this represents is estimated as $\sim1.8~\rm{years}$.

We then examined the differential distributions of predicted overall counts first from various single, and then combined populations of burst progenitor. By varying the parameters of the chosen luminosity functions, we compared the predicted overall counts ($dN/dp$) to the observed differential distribution from the $C_{max}/C_{min}$ table. For each set of LF parameters, the redshift distribution of sGRBs was calculated, and the nearby distributions compared with the observed correlated distributions from \citet{2005Natur.438..991T}. Note that we use an extended version of our previous correlation analysis out to 155~Mpc, and use the correlations measured against galaxies in concentric \textit{shells} (as opposed to \textit{spheres}) of recession velocity \citep[see][]{2007MNRAS.382L..21C} in order to obtain a local differential distribution for the model fitting. $\chi^{2}$ minimisation was then used to optimise the LF parameters by fitting simultaneously to the overall count rate and the local distributions. We assumed a Poissonian error distribution on the overall count rate, whereas we used the explicit Monte Carlo derived error distribution on the local correlated fraction (the error distributions from the Monte Carlo simulations closely follow a normal distribution, even at low correlation levels since the function defined in \citet{2005Natur.438..991T} is equally sensitive to anti-correlation giving rise to negative percentage correlations in those situations). Note that the greater number of data points in the number count fits means that the combined $\chi^2$ values are dominated by the goodness of fit to the count rate distribution. To explicitly ask whether any of our chosen single LFs can remain consistent with the BATSE number counts while being forced to produce a local distribution of bursts, we also find the best fits constrained by the correlated fraction alone.

In order to check the plausibility and consistency of the best fit models, we further compared the derived redshift distribution to that of sGRBs observed by \textit{Swift}. We caution that this sample is neither uniformly selected nor complete. Previous studies have analysed the early \textit{Swift} distributions~\citep[e.g.][]{2006A&A...453..823G,2006ApJ...643L..91H,2006ApJ...650..281N,2007arXiv0710.3099S,2008ApJ...675..566O}, and it is clearly useful to compare our models to the current best known redshift distribution in order to check that the predictions are not unrealistic. We stress that the \textit{Swift} distribution was not part of the statistical analysis. sGRB redshifts have so far only been found from host galaxy associations, the identification of which is not always unambiguous.  Furthermore, even the classification of some bursts as either short or long is controversial since their durations change substantially depending on whether or not emission from the long-soft tails (seen in a number of bursts) is included. Nevertheless, about a dozen probable short-hard bursts have reasonably secure redshifts. Specifically we include the following 10 sGRBs: GRBs 050509B, 050724, 051221a, 060801, 061006, 061201, 061210, 061217 (see \citet{2007ApJ...670.1254B} and references therein), 070714B~\citep{2007GCN..6836....1G} and 071227~\citep{2007GCN..7152....1G}. In order to produce the predicted \textit{Swift} redshift distribution, the \textit{Swift} BAT threshold for sGRBs was assumed to be twice that of BATSE~\citep{2006ApJ...644..378B}.
 
\section{Results}

\begin{table}
\caption{Results of single population Luminosity Functions, presented in order of decreasing overall goodness of fit (i.e. increasing overall $\chi^2/dof$). The LFs follow merger delay time (formation plus coalescence) distributions either \textit{flat} in $\log$ space ($dP_{m}/d(\log(t)) = constant $), or with a narrow \textit{spike} at early times, or the \textit{SMD} profile of Equation~\ref{mass}. The number of degrees of freedom ($dof$) for the $C_{max}/C_{min}$ and overall distributions are 22 and 26 respectively. $l_0$ is in units of $\log(\rm{erg~s^{-1}})$, $\sigma$ in dex and $\alpha$ is dimensionless.}
{\centering
\begin{tabular*}{1.0\linewidth}{ccccc}\hline\\
NS Merger&Parameters&Local&$C_{max}/C_{min}$&Overall\\
LF& ($l_0\equiv\log{L_0}$)&$\chi^2$&$\chi^2/dof$&$\chi^2/dof$\\
\hline\\
Schechter & $l_0=51.75$&11.93&1.01& 1.31 \\
(\textit{flat})     & $\alpha=1.25$ &        & \\
Schechter & $l_0=51.8$ &11.93&1.03& 1.33 \\
(\textit{spike})     & $\alpha=1.25$ &        & & \\
Schechter & $l_0=50.45$ &11.87&1.05& 1.34 \\
(\textit{SMD})     & $\alpha=0.9$ &        & & \\
Lognorm & $l_0=48.9$ &11.82&1.09& 1.38 \\
(\textit{SMD})   & $\sigma=0.75$ &        & &\\
Lognorm & $l_0=46.4$&11.88&1.18& 1.46 \\
(\textit{flat})   & $\sigma=1.5$ &        & &\\
Lognorm & $l_0=46.6$ &11.89&1.19& 1.47 \\
(\textit{spike})   & $\sigma=1.45$ &        & &\\
\hline\\
\end{tabular*}
}
\label{results_single}
\end{table}

\begin{figure}
  \includegraphics[width=1.0\linewidth]{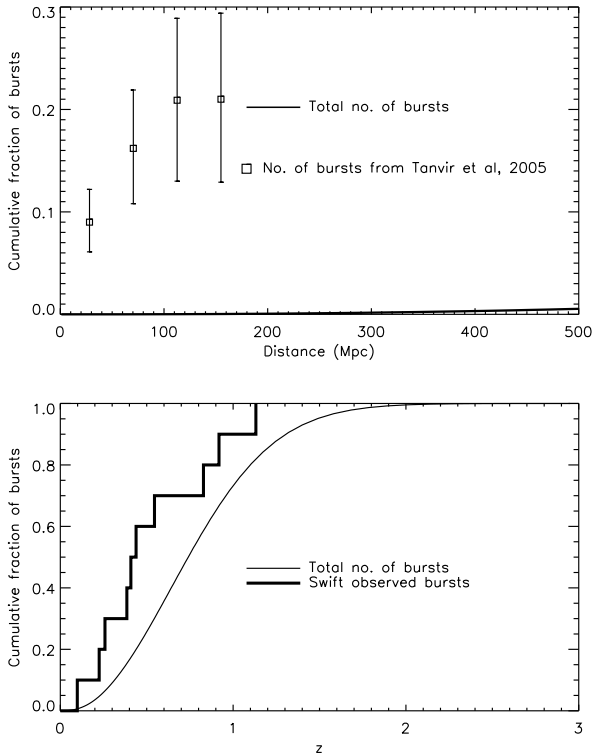}
  \caption{Burst distributions from the best fit merger single population Schechter function LF following a $dP_{m}/d(\log(t)) = constant $ merger time delay distribution. Top panel shows predicted sGRB distribution within 500~Mpc compared to the local burst fraction measured in \citet{2005Natur.438..991T}, bottom panel shows the predicted burst distribution out to $z=3$ normalised and compared to the \textit{Swift} distribution discussed in the text.
\label{schec}}
\end{figure}

Table~\ref{results_single} lists the best fit parameters found from fitting distributions produced by single population NS merger LFs simultaneously to both the overall number counts and the local population as described above. The table is ordered in decreasing overall goodness of fit (i.e. increasing combined $\chi^2/dof$). As mentioned previously, the combined $\chi^2$ is dominated by the fit to the overall BATSE number counts and, as expected, all our chosen single population LFs produce good fits to the $C_{max}/C_{min}$ data leading to acceptable overall fits as measured by the combined $\chi^2$. However, none of the single progenitor population LFs reproduce the local burst population expected from the correlation results: for example Figure~\ref{schec} shows the results from a single Schechter function LF which can be seen to produce effectively no sGRBs within 300~Mpc. To enable a quantitative comparison with later results, column 3 in Table~\ref{results_single} lists the $\chi^2$ results considering the fit to the four data points of the local distribution alone. Note that since the fit is constrained by the overall data set, the local $\chi^2$ value is quoted unreduced in this and all subsequent Tables since the precise number of degrees of freedom relevant to this subset alone is difficult to estimate formally.


In order to ascertain whether a single merger population can produce the local bursts and remain consistent with the $C_{max}/C_{min}$ data, we then fit single LF populations to the local distribution alone, with no constraints placed on goodness of fit to the overall number counts. As can be seen from Table~\ref{results_single_local}, single Schechter function LFs can produce a local population, but the associated number count distribution is an extremely poor match to the $C_{max}/C_{min}$ data.

\begin{table}
\caption{Results of single population Luminosity Functions constrained to fit the local distribution, presented in order of decreasing goodness of fit. Details as for Table~\ref{results_single}.}
{\centering
\begin{tabular*}{1.0\linewidth}{ccccc}\hline\\
NS Merger&Parameters&Local&$C_{max}/C_{min}$&Overall\\
LF& ($l_0\equiv\log{L_0}$)&$\chi^2$&$\chi^2/dof$&$\chi^2/dof$\\
\hline\\
Schechter & $l_0=51.1$&0.76&$>100$& $>100$ \\
(\textit{flat})     & $\alpha=2.2$ &        & \\
Schechter & $l_0=51.0$ &0.76&$>100$& $>100$ \\
(\textit{spike})     & $\alpha=2.2$ &        & & \\
Schechter & $l_0=53.0$ &1.08&$>100$& $>100$ \\
(\textit{SMD})     & $\alpha=2.15$ &        & & \\
Lognorm & $l_0=43.1$ &5.30&$>100$& $>100$ \\
(\textit{spike})   & $\sigma=1.35$ &        & &\\
Lognorm & $l_0=43.1$&5.38&$>100$& $>100$ \\
(\textit{flat})   & $\sigma=1.3$ &        & &\\
Lognorm & $l_0=43.0$ &6.43&$>100$& $>100$ \\
(\textit{SMD})   & $\sigma=1.4$ &        & &\\
\hline\\
\end{tabular*}
}
\label{results_single_local}
\end{table}

\begin{figure}
  \includegraphics[width=1.0\linewidth]{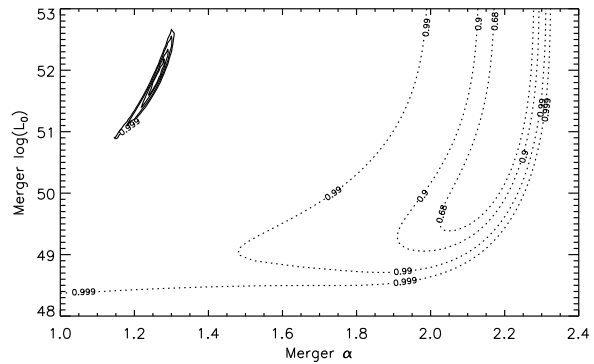}
  \caption{$\chi^2$ contours for the single population Schechter function LF ($dP_{m}/d(\log(t)) = constant $ merger delay time distribution) from Table~\ref{results_single}. The solid lines are the contours for the fits to the overall number counts, and the dotted lines are contours for the fits to the local distribution. Contours are plotted at 0.68, 0.9, 0.99 and 0.999 significance levels. For the sake of clarity, only the outermost 0.999 contour is labelled for the number count fit.\label{contourplot}}
\end{figure}

Of course, these results represent the best possible reproduction of the local bursts (in terms of minimum $\chi^2$ values), and it maybe that there exist poorer fits to the local population which are nevertheless better fits to the overall number counts. Figure~\ref{contourplot} shows the $\chi^2$ contours (individually for fits to both the overall number counts and the local distribution) for the single population Schechter function LF ($dP_{m}/d(\log(t)) = constant $ merger delay time distribution) from Table~\ref{results_single}. From the Figure it can be seen that acceptable fits to the overall number count distribution (to a significance level of 0.999) are found only in a narrow band of the LF parameter space, well separated from even the 0.99 significance level of the local distribution fits (which represents effectively no local distribution given the size of the errors on the correlation points shown in the top panel of Figure~\ref{schec} for example). Hence the possibility of this single Schechter function LF to reproduce the local distribution while remaining consistent with the overall number counts can be rejected with greater than $99.9\%$ confidence. Similar results with equivalent levels of rejection are found for the other single LF models.


The inability of the local constraint to produce a distribution which fits the number counts is effectively a consequence of the intrinsic sGRB rate calculated in Equation~\ref{rgrb} from the assumed merger rates: not enough bursts in total can be produced. The fit to the $C_{max}/C_{min}$ data can be improved to reasonable $\chi^2$ levels by increasing the intrinsic merger rate by a large factor ($\ge500$), but the overall redshift distribution produced as a consequence is extremely unrealistic, with all bursts produced within $z\sim0.1$.

\begin{figure}
  \includegraphics[width=1.0\linewidth]{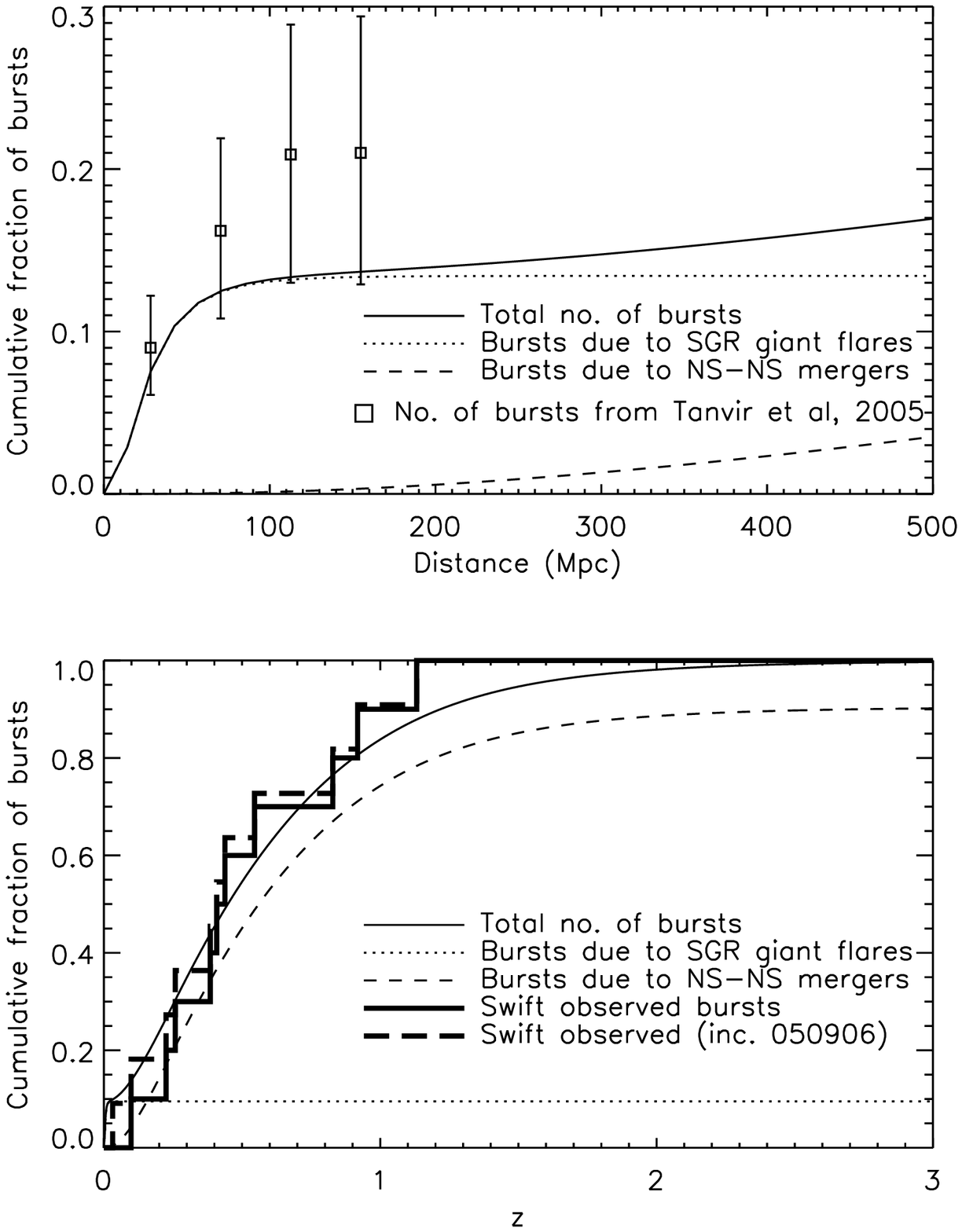}
  \caption{Burst distributions from dual lognormal LF (following $dP_{m}/d(\log(t)) = constant $ merger time delay distribution) populations. Panel Details as for Figure~\ref{schec}.
\label{lognorm}}
\end{figure}

\begin{figure}
  \includegraphics[width=1.0\linewidth]{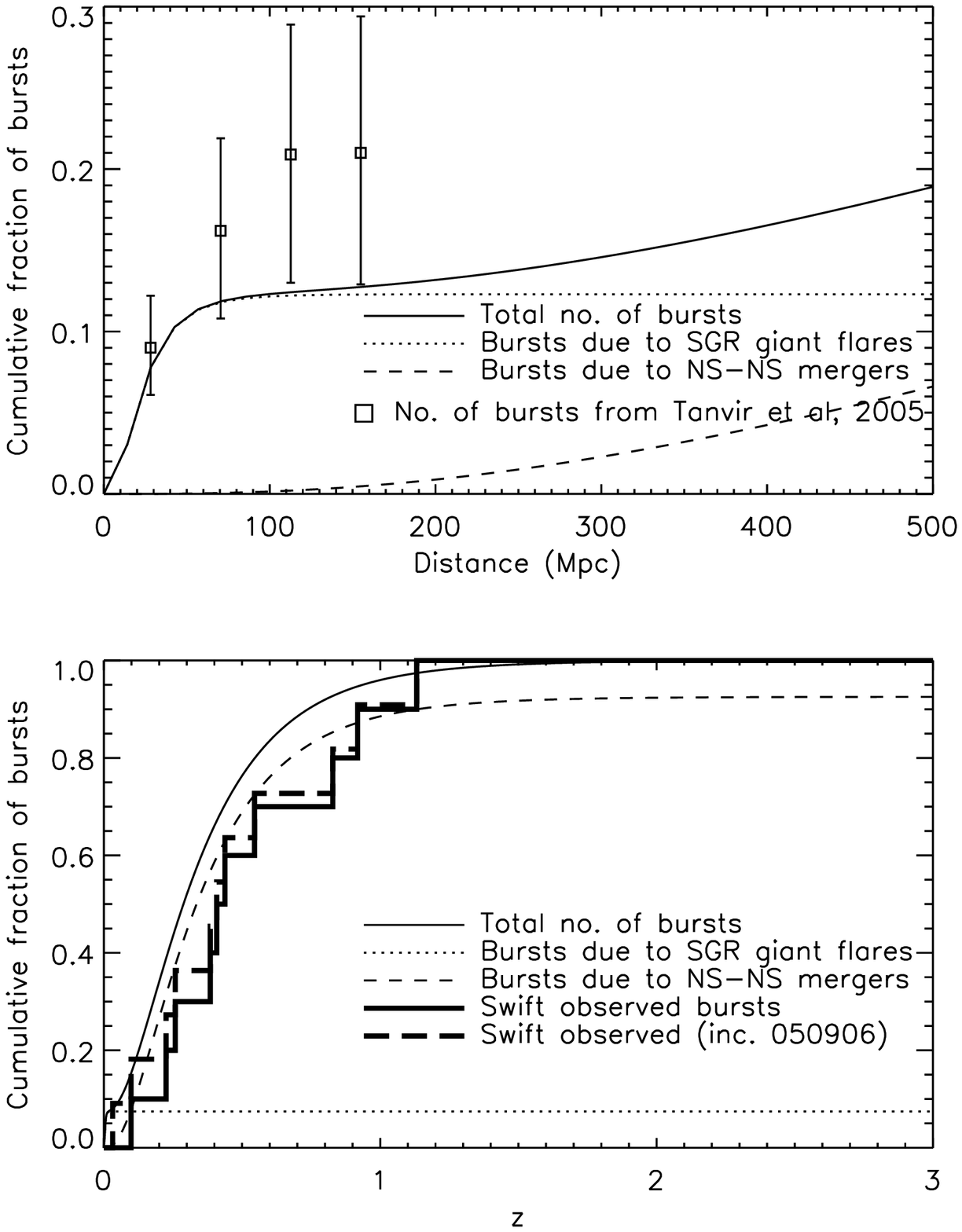}
  \caption{Burst distributions from dual lognormal LF (following SMD) populations. Panel Details as for Figure~\ref{schec}.\label{lognormsmd}}
\end{figure}

In contrast, Table~\ref{results} shows best fit LF parameters for various combinations of dual NS merger and SGR giant flare luminosity function models, along with their respective minimum $\chi^2$ values. As can be seen from Table~\ref{results_single_local}, and the examples of Figures~\ref{lognorm} and \ref{lognormsmd}, the local bursts are only ever reproduced by the lower luminosity LF. By minimising the combined $\chi^2$ values, all the dual LFs tested reproduced the local distribution well while retaining overall number count fits comparable to those of the single LFs. Furthermore, the best fit LF parameters of the dual models are reasonable, and the overall redshift distribution is much more realistic.

For example, a dual lognormal LF, with merger rates following either a delayed merger model (Figure~\ref{lognorm}) or the SMD model of Equation~\ref{mass} (Figure~\ref{lognormsmd}), produces a good fit to the expected local population while remaining consistent with the early \textit{Swift} redshift distribution. The upper panels of Figures~\ref{lognorm} and \ref{lognormsmd} show the comparison of these models to the local sGRB distribution determined by our BATSE cross-correlation analysis, and are typical in that all the dual populations reproduce this local population well. Since these data were used to constrain the fit, a good agreement is to be expected, but it is still interesting to note that the merger population contributes only a small fraction to these local bursts. The lower panels show the overall predicted redshift distribution.

\begin{table*}
\caption{Results of dual population Luminosity Functions, presented in order of decreasing goodness of fit (i.e. increasing overall $\chi^2/dof$). The LFs follow merger delay time (formation plus coalescence) distributions either \textit{flat} in $\log$ space ($dP_{m}/d(\log(t)) = constant $) or the \textit{SMD} profile of Equation~\ref{mass}. Also shown are two results normalised using order of magnitude different observed Galactic ($MW$) rates. The number of degrees of freedom ($dof$) for the $C_{max}/C_{min}$ and overall distributions are 20 and 24 respectively. $l_0$ is in units of $\log(\rm{erg~s^{-1}})$, $\sigma$ in dex and $\alpha$ is dimensionless.}
{\centering
\begin{tabular}{ccccccc}\hline\\
NS Merger&LF Parameters&SGR Giant Flare&LF Parameters&Local&$C_{max}/C_{min}$&Overall\\
LF& ($l_0\equiv\log{L_0}$)&LF& ($l_0\equiv\log{L_0}$)&$\chi^2$&$\chi^2/dof$&$\chi^2/dof$\\
\hline\\
Schechter & $l_0=52.3$ & Power law & $l_0=46.7$ &2.03&1.15& 1.04\\
(\textit{flat})    & $\alpha=1.3$ &(\textit{flat})    &  $\alpha=1.25$ &&&\\
Schechter & $l_0=52.3$ & Lognorm & $l_0=45.2$ &1.45&1.20& 1.06 \\
(\textit{flat})    & $\alpha=1.3$ &(\textit{flat})  &  $\sigma=0.6$ &&&\\
Lognorm & $l_0=48.35$ & Lognorm & $l_0=45.3$ &1.66&1.31&1.16 \\
(\textit{SMD})    & $\sigma=1.05$ &(\textit{SMD})  &  $\sigma=0.55$ &&&\\
Lognorm & $l_0=47.2$ & Power law & $l_0=46.7$ &2.06&1.69&1.49 \\
(\textit{flat})  & $\sigma=1.2$ &(\textit{flat})  &  $\alpha=1.35$ &&&\\
Lognorm & $l_0=47.05$ & Lognorm & $l_0=45.2$ &1.55&1.72&1.50 \\
(\textit{flat})  & $\sigma=1.2$ &(\textit{flat}) &  $\sigma=0.6$ &&&\\
\hline\\
Lognorm & $l_0=48.6$ & Lognorm & $l_0=44.1$ &1.57&1.36&1.20 \\
(\textit{SMD})    & $\sigma=0.9$ &(\textit{SMD})  &  $\sigma=0.8$ &&&\\
& &($10 \times MW $) & &&&\\
Lognorm & $l_0=48.6$ & Lognorm & $l_0=46.3$ &3.13&1.28&1.20\\
(\textit{SMD})    & $\sigma=0.9$ &(\textit{SMD})  &  $\sigma=0.2$ &&&\\
& &($0.1 \times MW $) & &&&\\
\hline
\end{tabular}
}
\label{results}
\end{table*}

As mentioned before, the intrinsic Galactic rates used to normalise the LFs are not well constrained. Hence in Table~\ref{results} we also show the results of varying the intrinsic SGR flare rate up and down by an order of magnitude for the dual lognorm (SMD) fit of Figure~\ref{lognormsmd}. The production of a local sGRB population is robust against this change, and the overall fit remains good. As may be expected, an increase in the intrinsic flare rate leads to the best fit SGR LF being moved down in luminosity, thus removing a greater fraction of the total flares from observability. Likewise, a lower intrinsic rate generates a higher (and narrower) LF distribution, though in both cases the LF parameters remain entirely plausible.

\begin{figure}
  \includegraphics[width=1.0\linewidth]{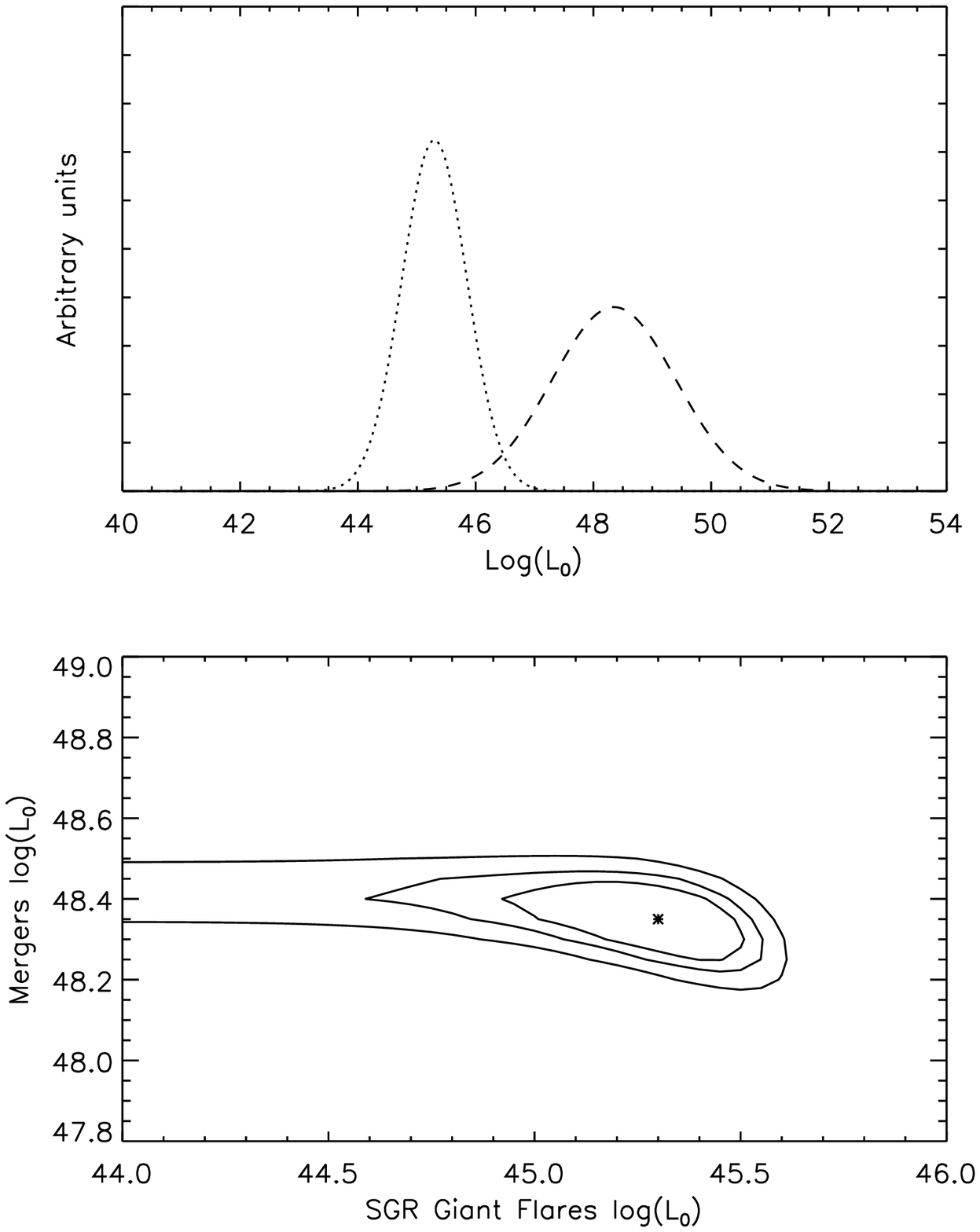}
  \caption{Best fit dual population LFs from Figure~\ref{lognormsmd}. The LFs (top panel: dotted line SGR giant flares, dashed line mergers) are lognormal with intrinsic merger rate components following the SMD model of Equation~\ref{mass}. The bottom panel shows contours of $\chi^2$ in $\log(L_0)$ space. Contours shown represent 0.68, 0.9 and 0.99 confidence limits with the minimum $\chi^2$ value plotted as an asterisk. \label{contours}}
\end{figure}

Figure~\ref{contours} shows the best fit LFs and associated contours of $\chi^2$ with respect to $L_0$ for the dual population from Figure~\ref{lognormsmd}. Despite the uncertainties in the underlying Galactic rates of the models, the best fit parameters obtained for this and the other dual LFs are plausible given the known properties of SGR giant flares and classic sGRB luminosities. We note that the slopes of the SGR flare power law LFs obtained (1.25 - 1.35) are shallower than the slopes found for ordinary SGR burst fluence distributions (1.4 - 1.8)~\citep{1996Natur.382..518C,2000ApJ...532L.121G}.

\section{Discussion}
 
The lower panels of Figures~\ref{lognorm} and \ref{lognormsmd} imply that \textit{Swift} should have triggered on about one SGR flare to date (this would rise by a factor of $\sim2$ if the redshift completeness for such flares were greater than for sGRBs as a whole, as is likely given that low-redshift host galaxies are easily identified). We note that a possible candidate is GRB\,050906, which may have originated in a galaxy at $\approx130$~Mpc~\citep{2008MNRAS.384..541L}, and the preliminary \textit{Swift} redshift distributions in Figures~\ref{lognorm} and \ref{lognormsmd} are plotted both including and excluding this burst.

There are two further recent sGRB events which are candidate extragalactic SGR flares, though neither triggered \textit{Swift}: GRB\,051103 whose IPN error box includes
the outskirts of M81 at 3.5~Mpc~\citep{2005GCN..4197....1G}, and GRB\,070201 whose error box similarly overlaps a spiral arm of M31 at only $\sim0.77$~Mpc~\citep{2007GCN..6091....1P,2007GCN..6098....1P,2007arXiv0712.1502M}. Both have characteristics
of SGR giant flares~\citep{2007AstL...33...19F,2007arXiv0712.1502M,2007arXiv0712.3585O}, and furthermore the non-detection of gravitational waves by LIGO from GRB\,070201~\citep{2007arXiv0711.1163L} excludes a merger progenitor within M31 with $>99\%$ confidence. If both these events were due to extragalactic SGRs then this brings to three the number of giant flares with peak luminosity $>10^{47}\rm{erg~s^{-1}}$ seen in just a few years.

\citet{2008MNRAS.384..541L} estimated that a Galactic SGR giant flare rate of $\sim0.5\times10^{-4}~\rm{yr^{-1}}$ would be sufficient to produce $\sim10$ extragalactic flares within a sphere of radius 100 Mpc. Using a power law LF (constrained by a search for positional coincidences between galaxies within 20~Mpc and the IPN error boxes of a sample of 47 sGRBs), \citet{2007ApJ...659..339O} estimated the rate of extragalactic flares with energy $>3.7\times10^{46}$erg (the energy of the 2004 SGR1806-20 event~\citep{2005Natur.434.1098H}) to be $\sim0.5\times10^{-4}~\rm{yr^{-1}}$ per SGR, and the $95\%$ confidence lower limit of the Galactic rate to be $2\times10^{-4}~\rm{yr^{-1}}$ per SGR. Our analysis estimates the rate of flares with peak luminosity $>10^{47}~\rm{erg~s^{-1}}$ to be between these two values at $\sim1\times10^{-4}\rm~{yr^{-1}}$ per SGR. We estimate the SFR of galaxies within 5~Mpc listed by \citet{2007ApJ...659..339O} (with revised distance estimates~\citep{2004AJ....127.2031K}) to be about $22\times$ that of the Milky Way. Adopting our predicted (lognorm following SMD) flare rate, the probability of observing two (one) or more such flares within this volume during the 17 years of IPN3 observation is $1\%~(14\%)$. This indicates we have been witness to a rather rare coincidence, and is perhaps suggestive that not both GRB\,051103 and GRB\,070201 are SGR flares.

\section{Summary and Conclusions}

We have examined a selection of plausible Luminosity Functions, singly and in combination, for both neutron star mergers and SGR giant flares as progenitors of short Gamma-ray Bursts. Assuming observed and theoretical Galactic intrinsic rates, merger delay time distributions, Star Formation Rate and Stellar Mass Density parameterisations, we exclude both lognormal and Schechter type LFs for a single NS merger population of progenitor as being unable to produce a nearby sGRB population while remaining consistent with overall BATSE number counts. Indeed, given that even a Schechter function (dominated by low luminosity events) cannot reproduce the likely local population, it is hard to conceive of any unimodal LF which could and still be consistent with the higher redshift distribution. We suggest that at least a bimodal LF, and therefore likely a dual population model, is necessary to account for the local population. Given the uncertainties in the intrinsic rates assumed, we cannot sensibly choose between the LF combinations, but we point out that the best fit LF parameters in all dual populations considered are in reasonable agreement with the known properties of SGR giant flares and classic sGRBs, even when the intrinsic rate of Galactic SGR flares is varied by an order of magnitude in either direction. To put this another way, as is well known a single population Luminosity Function provides a good fit to overall BATSE number counts, but we find that a separate, lower luminosity population of progenitors is both required, and is sufficient, to reproduce a local sGRB population. Furthermore, the properties of this population are in agreement with those observed from Galactic SGR giant flares.


\section*{Acknowledgements}
We thank P\'all Jakobsson and Andrew Levan for useful discussions. RC and RSP acknowledge the support of the University of Hertfordshire. NRT acknowledges the support of a UK STFC senior research fellowship. We are grateful to the anonymous referee, whose attention to detail significantly enhanced the paper.

\bibliographystyle{mn2e_281004}
\bibliography{bobrefs}
\clearpage
\appendix
\section{The low luminosity cutoff, $L_{min}$, in the Schechter Luminosity Functions}

\begin{figure}
  \includegraphics[width=1.0\linewidth]{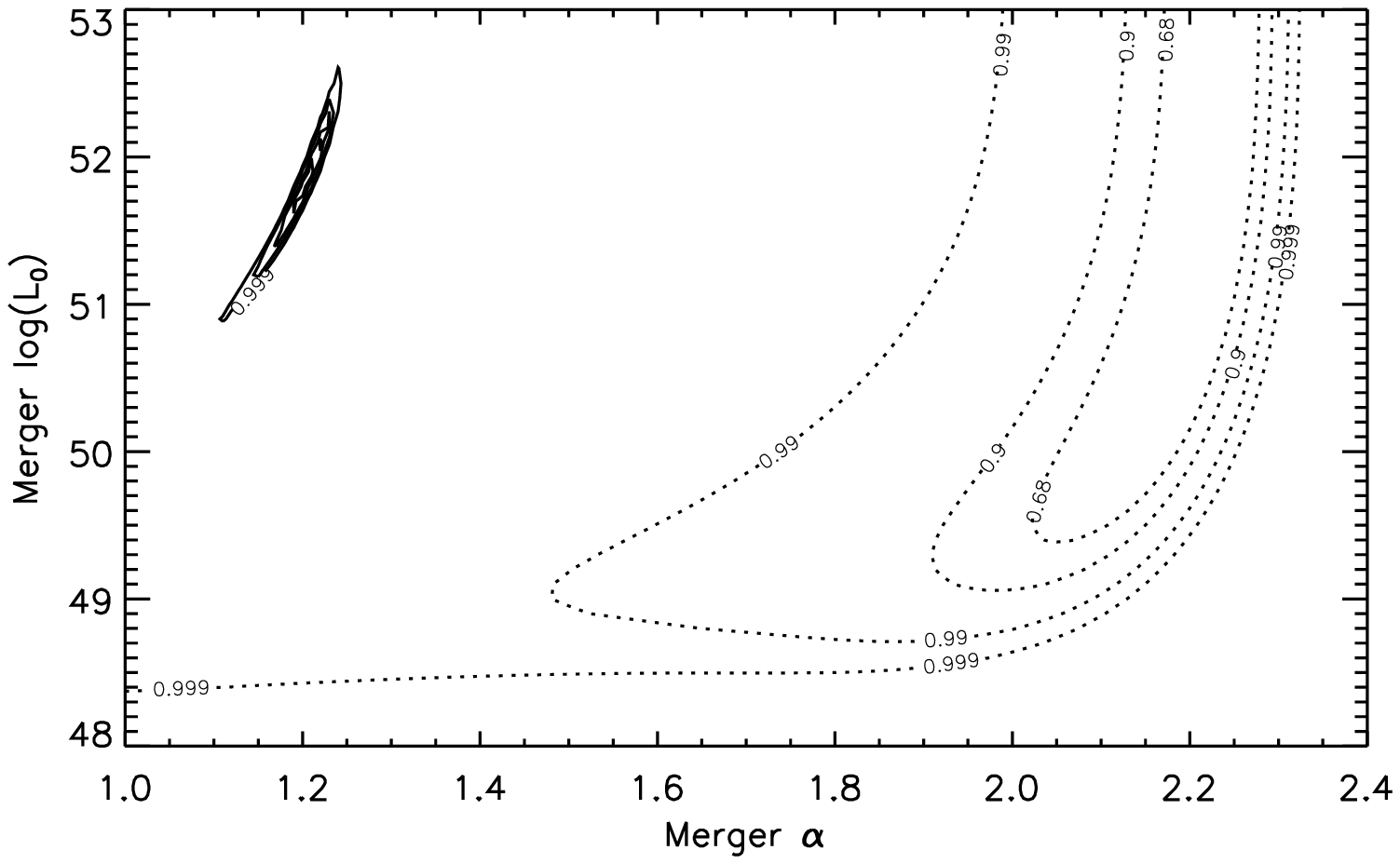}
  \caption{$\chi^2$ contours for the single population Schechter function LF as in Figure~\ref{contours} of the main text,  with $L_{min}=10^{40}~\rm{erg~s}^{-1}$. Details as for Figure~\ref{contours} of main text.\label{contourplot40}}
\end{figure}

\begin{figure}
  \includegraphics[width=1.0\linewidth]{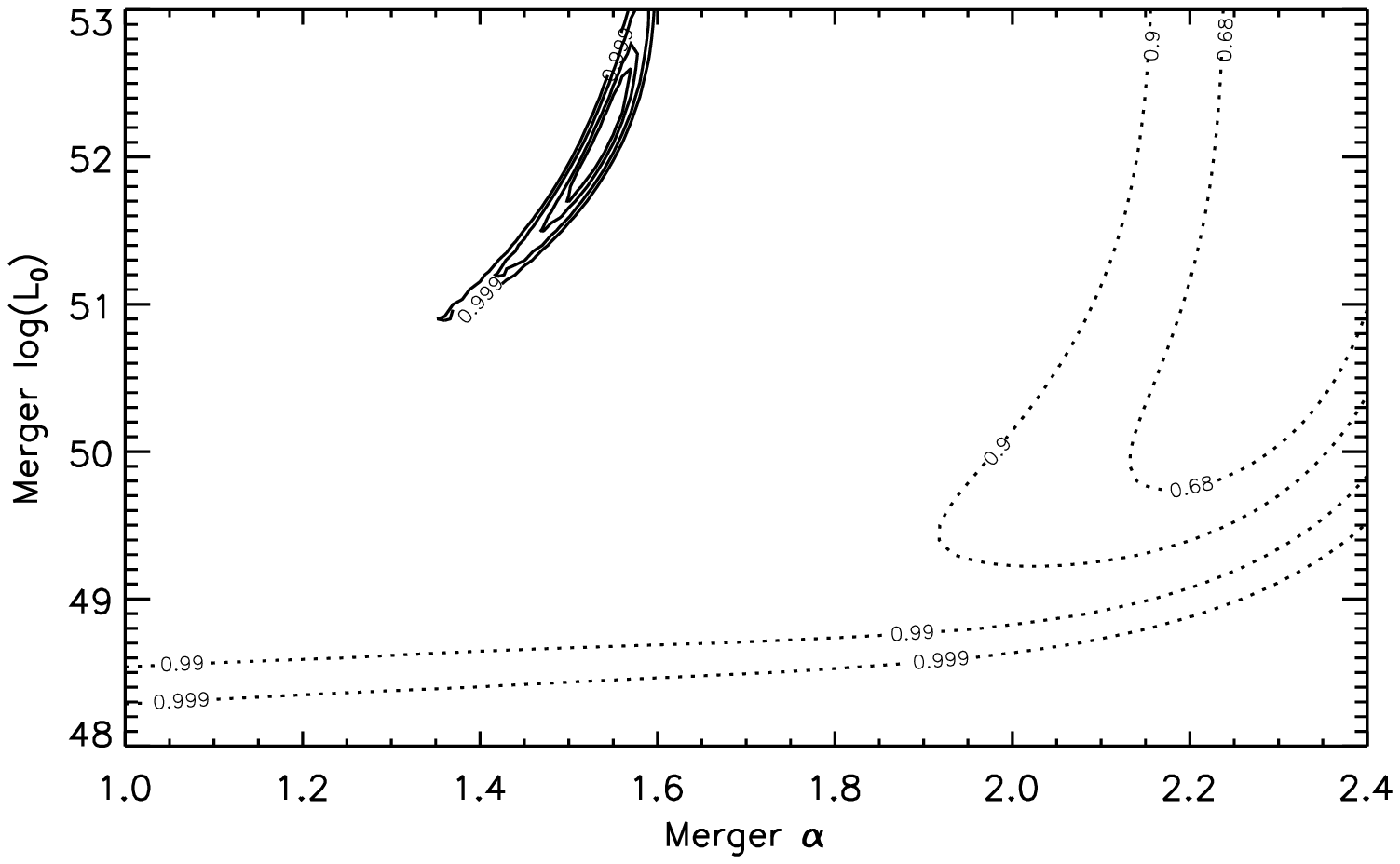}
  \caption{$\chi^2$ contours for the single population Schechter function LF as in Figure~\ref{contours} of the main text,  with $L_{min}=10^{46}~\rm{erg~s}^{-1}$. Details as for Figure~\ref{contours} of main text.\label{contourplot46}}
\end{figure}

It may be thought that the precise choice of the lower luminosity cutoff, $L_{min}$, necessary for convergence of the Schechter type Luminosity Functions has a significant effect on the LF parameters found, and further on the ability of that LF to reproduce the local population while remaining consistent with the overall number counts. We therefore tested a range of low luminosity cutoffs ($L_{min}=10^{40}~\rm{erg~s}^{-1}$ to $10^{46}~\rm{erg~s}^{-1}$, and found that the best fit LF parameters are relatively insensitive to the chosen $L_{min}$ as shown in Table~\ref{lmintab}. Furthermore, the separation in LF parameter space is maintained between the best fits to the number counts and the best fits to the local distribution regardless of choice of $L_{min}$ as shown for example in Figures~\ref{contourplot40} and \ref{contourplot46}. 
\newpage
\begin{table}
\caption{Results of single population Schechter Luminosity Functions with varying values of $L_{min}$. Details as for Table~\ref{results_single} of main text.}
{\centering
\begin{tabular*}{1.0\linewidth}{ccccc}\hline\\
$\log(L_{min})$&LF Parameters&Local&$C_{max}/C_{min}$&Overall\\
LF& ($l_0\equiv\log{L_0}$)&$\chi^2$&$\chi^2/dof$&$\chi^2/dof$\\
\hline\\
40& $l_0=51.80$&11.93&$0.99$&1.30\\
& $\alpha=1.20$ &        & \\
42& $l_0=51.75$ &11.93&$1.01$&1.33\\
& $\alpha=1.25$ &        & & \\
44& $l_0=52.00$ &11.92&$1.00$&1.30\\
& $\alpha=1.36$ &        & & \\
46& $l_0=52.00$ &11.89&$1.04$&1.34\\
& $\sigma=1.53$ &        & &\\
\hline\\
\end{tabular*}
}
\label{lmintab}
\end{table}

\label{lastpage}

\end{document}